\documentclass[sigconf]{acmart}
\usepackage{comment}
\usepackage{subcaption}
\usepackage{multirow}
\usepackage{adjustbox}
\usepackage{array,multirow,graphicx}
\usepackage{float}
\AtBeginDocument{%
  \providecommand\BibTeX{{%
    \normalfont B\kern-0.5em{\scshape i\kern-0.25em b}\kern-0.8em\TeX}}}

\begin{document}
\acmYear{2021}\copyrightyear{2021}
\setcopyright{acmcopyright}
\acmConference[WiseML '21]{3rd ACM Workshop on Wireless Security and Machine Learning}{June 28--July 2, 2021}{Abu Dhabi, United Arab Emirates}
\acmBooktitle{3rd ACM Workshop on Wireless Security and Machine Learning (WiseML '21), June 28--July 2, 2021, Abu Dhabi, United Arab Emirates}
\acmPrice{15.00}
\acmDOI{10.1145/3468218.3469041}
\acmISBN{978-1-4503-8561-9/21/06}
\title{Intermittent Jamming against Telemetry and Telecommand of\\ Satellite Systems and A Learning-driven Detection Strategy}

\author{Selen Gecgel}
\authornote{Corresponding author.\footnotemark[1] \footnotemark[2]}
\authornote{Department of Electronic Engineering, Turkish Air Force Academy, National Defense University, Istanbul, Turkey}
 \orcid{}
 \affiliation{%
 	\institution{Istanbul Technical University\footnotemark[1]\\ Wireless Communication Research Laboratory}
 	\city{Istanbul}
 	\country{Turkey}}
 \email{gecgel16@itu.edu.tr}
 \affiliation{%
 	\institution{Turkish Air Force Academy\footnotemark[2]\\National Defense University}
 	\city{Istanbul}
 	\country{Turkey}}
 \email{sgecgel@hho.edu.tr}
\author{Gunes Karabulut Kurt\footnotemark[1]}
\authornote{Department of Electrical Engineering, Polytechnique Montreal, Monteal, QC, Canada (e-mail: gunes.kurt@polymtl.ca)}
 \affiliation{%
 	\institution{Istanbul Technical University\footnotemark[1] \\ Wireless Communication Research Laboratory}
 	\city{Istanbul}
 	\country{Turkey}}
 \email{gkurt@itu.edu.tr}
 \affiliation{%
 	\institution{Polytechnique Montreal\footnotemark[3]\\Department of Electrical Engineering}
 	\city{Monteal, QC}
 	\country{Canada}}
 \email{gunes.kurt@polymtl.ca}
\begin{abstract}
Towards sixth-generation networks (6G), satellite communication systems, especially based on Low Earth Orbit (LEO) networks, become promising due to their unique and comprehensive capabilities. These advantages are accompanied by a variety of challenges such as security vulnerabilities, management of hybrid systems, and high mobility. In this paper, firstly, a security deficiency in the physical layer is addressed with a conceptual framework, considering the cyber-physical nature of the satellite systems, highlighting the potential attacks. Secondly, a learning-driven detection scheme is proposed, and the lightweight convolutional neural network (CNN) is designed. The performance of the designed CNN architecture is compared with a prevalent machine learning algorithm, support vector machine (SVM). The results show that deficiency attacks against the satellite systems can be detected by employing the proposed scheme.
\end{abstract}

\maketitle
\section{Introduction} 
The foresight becomes prevalent that satellite communication systems are the major parts of next-generation wireless networks due to ubiquitous connectivity and coverage opportunities \cite{int4}. The main drawbacks of satellite communications are propagation delay and packet loss. However, Low Earth Orbit (LEO) satellites alleviate these problems to a reasonable level. Additionally, LEO satellites come into prominence with lower launching and production costs compared to high-altitude satellites. Therefore, their seamless connectivity with terrestrial networks are envisioned to extend the globalization and quality of mobile services, as new LEO constellations have emerged with an increasing satellite number such as OneWeb, Telesat, and Starlink \cite{int1,int2,int3}. Even though they offer numerous advantages, LEO satellite networks bring significant security requirements and impending issues. 

\begin{figure*}[ht!]
  \includegraphics[width=0.95\textwidth]{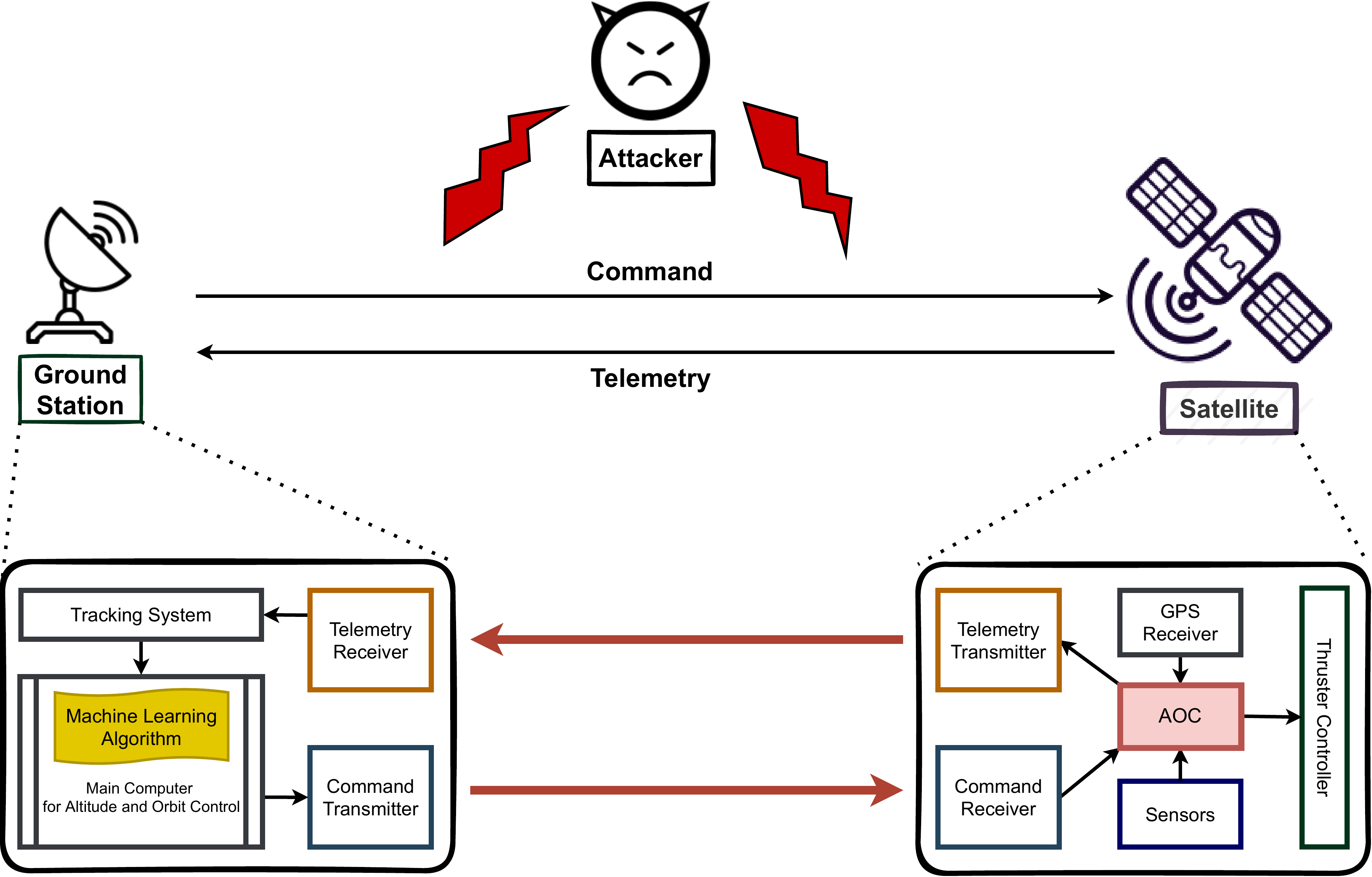}
  \caption{An illustration of the learning-driven detection scheme against physical layer attacks for telemetry and telecommand information.}
  \label{fig1}
\end{figure*}
Security concerns in satellite communication systems have been gaining increasing attention, and emerging research studies offer new solutions and analyses for different scenarios. For instance, the authors of \cite{int7} provide an investigation of the physical layer security for multi-user satellite communication and provide a threshold-based scheduling scheme. Cooperative and non-cooperative beamforming schemes are proposed to maximize the secrecy rate in \cite{int8}, where satellite and terrestrial base stations are operated at millimeter-wave frequencies. However, all of these studies focus on the legitimate user data security and miss the attacks against space-mission data systems.

Maintaining the correct orbit and altitude is one of the most critical points for continuous communication \cite{int11}. The satellites in LEO are exposed to a more substantial gravity impact by the Earth than high-altitude satellites. Therefore, they require an altitude and orbit control (AOC) system to provide stabilization. As summarized in Figure \ref{fig1}, the AOC system takes location data from the GPS receiver and sensors, and it can command maneuver. Actually, maneuver decisions are mainly given by the telemetry, tracking, command, and monitoring system (TTC\&M) which is located the ground station. TTC\&M systems sustain operational managements of satellites by conveying telemetry and command signals. An attack on telemetry or command signals can lead to interruptions in the communication services of LEO satellites or collisions of satellites. These attacks can be addressed in several aspects, such as confidentiality, authentication, data integrity, access control, or availability \cite{int5}. In this study, the threats against the telemetry or command data availability are tackled in the physical layer by considering jamming attacks. 

The presence of jamming attacks must be detected primarily to avoid or prevent jamming attacks. In \cite{int9}, a generalized likelihood ratio test based jammer detection is proposed by utilizing unused pilots. In \cite{int10}, two detectors based on variance and standard variance normalization are presented against jamming attack for legitimate communication. The authors of \cite{int12} propose a smart jamming detection method depending on the time-series analysis. However, these solutions may not be sufficient to detect some special attacks. In the recent years, solutions based on machine learning algorithms are widely used for jamming attack detection and generation \cite{shi1,shi2}. In \cite{int13}, the authors present a learning-driven detection methodology based on support vector machine (SVM) and reduce the hardware complexity. In \cite{int14}, machine learning algorithms are utilized to learn spectrum and determine transmit decisions for both the transmitter and the jammer. The authors proposed learning-driven jamming identification schemes in \cite{datac1} and \cite{datac2} by employing convolutional neural networks (CNN), recurrent neural networks, and SVM. They show attractive performance thanks to time-frequency transformation based data pre-processing. 

In this study, we firstly address a critical security vulnerability for LEO satellite networks that may vary depending on the frame contents, in accordance with the telemetry and telecommand signaling structure. Motivated by this problem, three different jamming attacks are considered by employing a conceptual signal structure; barrage jamming, pilot jamming, and intermittent jamming. A jammer detection methodology is proposed based on shallow and deep learning algorithms. The results prove that both detector schemes provide high performance. However, the proposed CNN-based detection scheme offers a superior performance for high SNR compared to its SVM-based counterpart. 

The rest of this study is organized as follows. In Section \ref{II}, a system model is introduced conceptually to generate telemetry and telecommand signals. The attacks are presented by considering the target security concern. In Section \ref{III}, the learning-driven detection schemes based on the proposed CNN architecture and SVM benchmark algorithm are provided. The numerical results are discussed in Section \ref{IV}. Finally, in Section \ref{vv}, the conclusions are drawn.
\section{System Model}\label{II}
\begin{figure*}[!h]
  \includegraphics[width=\linewidth]{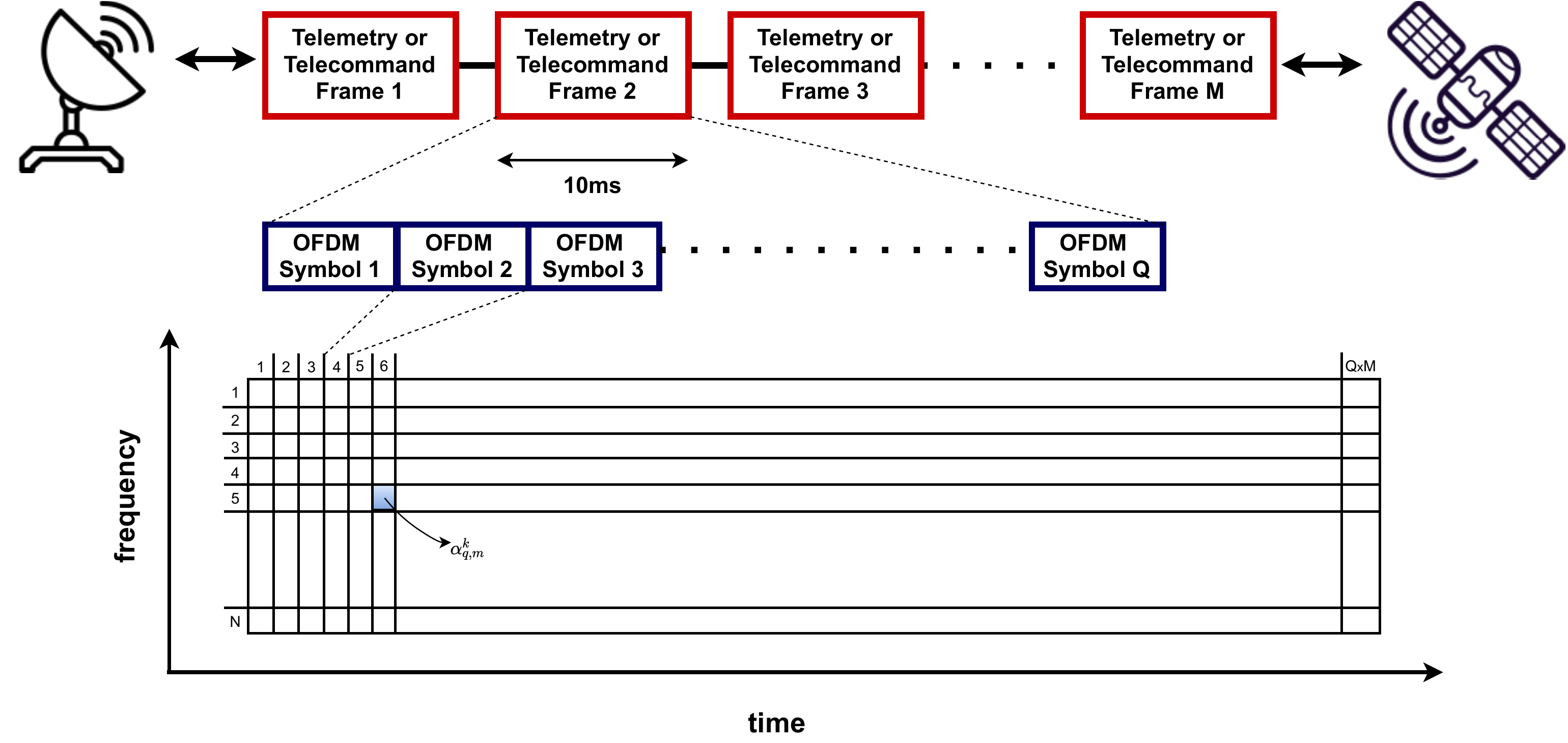}
  \caption{A conceptual telemetry and telecommand signal structure.}
  \label{fig2}
\end{figure*}
\subsection{A Conceptual Telemetry and Telecommand Signal Structure}

In this paper, an orthogonal frequency division multiplexing (OFDM) based satellite communication system is considered \cite{int6, OFDM}. Therefore, telemetry and telecommand signals are structured conceptually as $M$ temporal OFDM frames in Figure \ref{fig2}. Each frame is delivered in $10$ milliseconds and composed of $Q$ symbols. $N$ denotes the total number of subcarriers. An OFDM unit, $\alpha_{q,m}^{k}$, is located in the $q^\mathrm{{th}}$ symbol of the $m^\mathrm{{th}}$ frame in the time-frequency mapping of the signals, and it includes ${X(k)}$ as the $k^\mathrm{{th}}$ subcarrier.

Signal generation based on the aforementioned structure is realized by following the steps. The binary information is modulated using BPSK. Pilot symbols are inserted uniformly between subcarriers for the estimation of the channel state information. They are located based on comb-type pilot arrangement as
\begin{equation}\label{rsjeqm}
X(k)= \left\{\begin{matrix}
X_p(r) \ , & \bmod{(k,l)} = 4,  \\ 
0, & \ \ \text{otherwise} 
\end{matrix}\right. 
\end{equation}
where $r = 0, 1, \cdots ,(N_p-1)$, ${N_p}$ is the number of pilots, $\mathbf{X_p} = \left[X_p(0), X_p(1), \! \cdots, \!  X_p(N_p-1)\right]^{T}$ is the pilot sequence, and $l$ is the interval between consecutive pilot symbols. Null-carriers are inserted to both sides of modulated subcarriers, and the inverse fast Fourier transform (IFFT) operation is employed to the data sequence. It is transformed as
\begin{equation}
    x(n) = IFFT \left \{ X(k) \right \}.
\end{equation}
Following this step, the guard interval is appended to prevent inter-symbol interference. By considering line-of-sight propagations, Rician channel model is used \cite{Rician}. The received signal is defined as 
\begin{equation}
    y(n) = x(n) \otimes h_1(n) + w(n),
\end{equation}
where $\otimes$ is the convolution operator and $h_1(n)$ is the impulse response of the channel between the transmitter and the receiver. $w(n)$ is an independent and identically distributed (i.i.d.) zero-mean additive white Gaussian noise with the distribution of $\mathcal{CN}(0,\,\sigma_n^{2})$.

\subsection{Attacker Models}
\begin{figure*}[!h]
  \includegraphics[width=\textwidth]{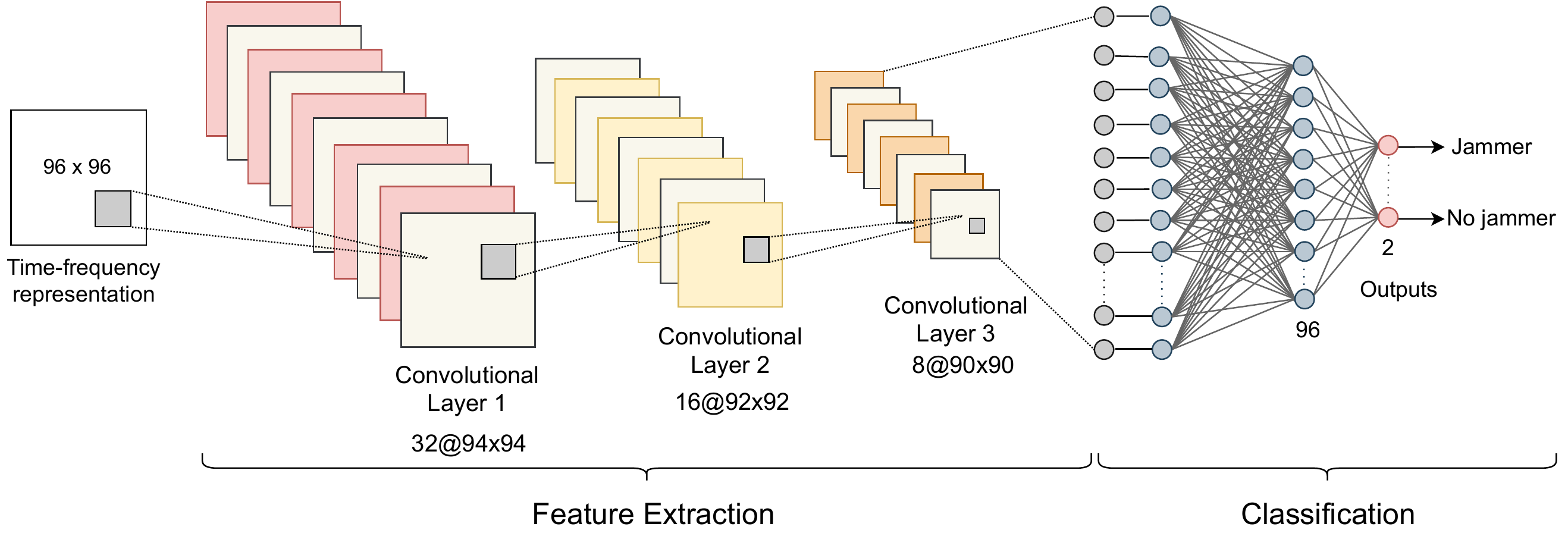}
  \caption{The proposed CNN architecture.}
  \label{cnn}
\end{figure*}
This study considers three different types of jamming strategies: barrage jamming, pilot tone jamming, and intermittent jamming. The jammed signal in the frequency domain in matrix form can be written as
\begin{equation}\label{received}
    \mathbf{Y}= \mathbf{H_1}\mathbf{X} + \mathbf{W} + \mathbf{H_{2}J},
\end{equation}
where $\mathbf{X} = \bigl[X(0),\ X(1),\ \cdots,\ X(k),\ \cdots,\ X(N-1)\bigl]^{T}$ and $ \mathbf{Y}=\bigl[Y(0), \ Y(1),\ \cdots, Y(k),\ \cdots,\ Y(N-1)\bigl]^{T}$ are the transmitted and received symbol sequences. $\mathbf{J} = \bigl[J(0),\ J(1),\ \cdots,\ J(k), \cdots,\ J(N-1)\bigl]^{T}$ denotes the jamming signal, and $\mathbf{H_2}$ is the channel matrix between the jammer and receiver. The following jamming attacks are considered.

\begin{itemize}
    \item \textbf{Barrage Jamming Attack:} When prior knowledge about the target signal does not exist, barrage jamming is optimal jamming attack \cite{barrage}. It can be defined simply as a noise attack to the whole transmission bandwidth. The received signal in (\ref{received}) under this type of attack is defined with the jamming signal vector $\mathbf{J}$ that is a zero-mean white Gaussian noise with the distribution of $\mathcal{CN}(0,\,\sigma^{2})$.
    
     \item \textbf{Pilot Tone Jamming Attack:} The pilot tone jamming attack is a power-efficient jamming strategy compared to barrage jamming. Only pilot tone frequencies are subjected to noise attacks; therefore, it depends on the existence of the knowledge about pilot symbol locations.
     
     \item \textbf{Intermittent Jamming Attack:} The intermittent jamming attack is designed by taking into account telemetry signal contents, including various information about satellite systems such as satellite position, sensor data, and subsystem conditions. If the attacker is aware of telemetry signal content, they do not continuously attack the time or frequency domain. We assume that the adversary knows target information patterns in telemetry data and attacks by following this pattern. Intermittent jamming signals, $\mathbf{J_I} = \bigl[J(0),\ J(1),\ \cdots, \ J(N-1)\bigl]^{T}$, are generated with the distribution of $ \mathbf{J_I} \sim  \mathcal{CN}(0,\,\sigma^{2})$ by employing pattern as
     \begin{equation}
         J(k)= \left\{\begin{matrix} 
         J(i) \ , & \bmod{(k,l)} = A_p \ and \  \bmod{(m,c)} =  d \\ 0, & \ \ \text{otherwise} \end{matrix}\right. 
     \end{equation}
     where $i = 0, 1, \cdots ,(N-1)$, and $c$ represents the interval between the OFDM symbols under this type of attack. $A_p$ and $d$ denote the attack periods in frequency and time. For the numerical analysis, we have considered $A_p= 4$ and $d=5$.
\end{itemize}

\section{Learning-driven Detection}\label{III}

In this study, we propose a learning-driven detection scheme by employing suitable machine learning algorithms. The detection problem is defined as a binary classification problem. Two classifiers, one based on SVM, and the other on the proposed CNN scheme, are compared.

\subsection{The Proposed CNN Architecture}
The CNN solution is a reasonable choice in classification tasks because of its adaptability, reusability, and operation portability. Convolutional layers will behave as a configurable feature extractor thanks to the convolutional filtering property. Essential components in the CNN solution are the data quality and data quantity. Data quality has an actual impact on the result of feature extraction. Proper normalization and processing should be applied to the data before being fed to the CNN. Data quantity is another crucial point since less number of samples causes over-fitting. Regularization tools help to overcome the over-fitting effect. Batch normalization and dropout layers are employed to reflect the normalization effect in the training process. 

The layers are employed consecutively in Figure \ref{cnn}. Convolutional kernels are initialized randomly to generate uniform distribution. In inter-layers ReLU activation function is utilized to take advantage of simplicity. However, in the last layer softmax activation function is employed for its suitability to solve classification problems. The training process is performed by using the Adam optimizer algorithm \cite{ADAM} and categorical-cross entropy loss function.

\begin{figure}[!h]
  \includegraphics[width=\linewidth]{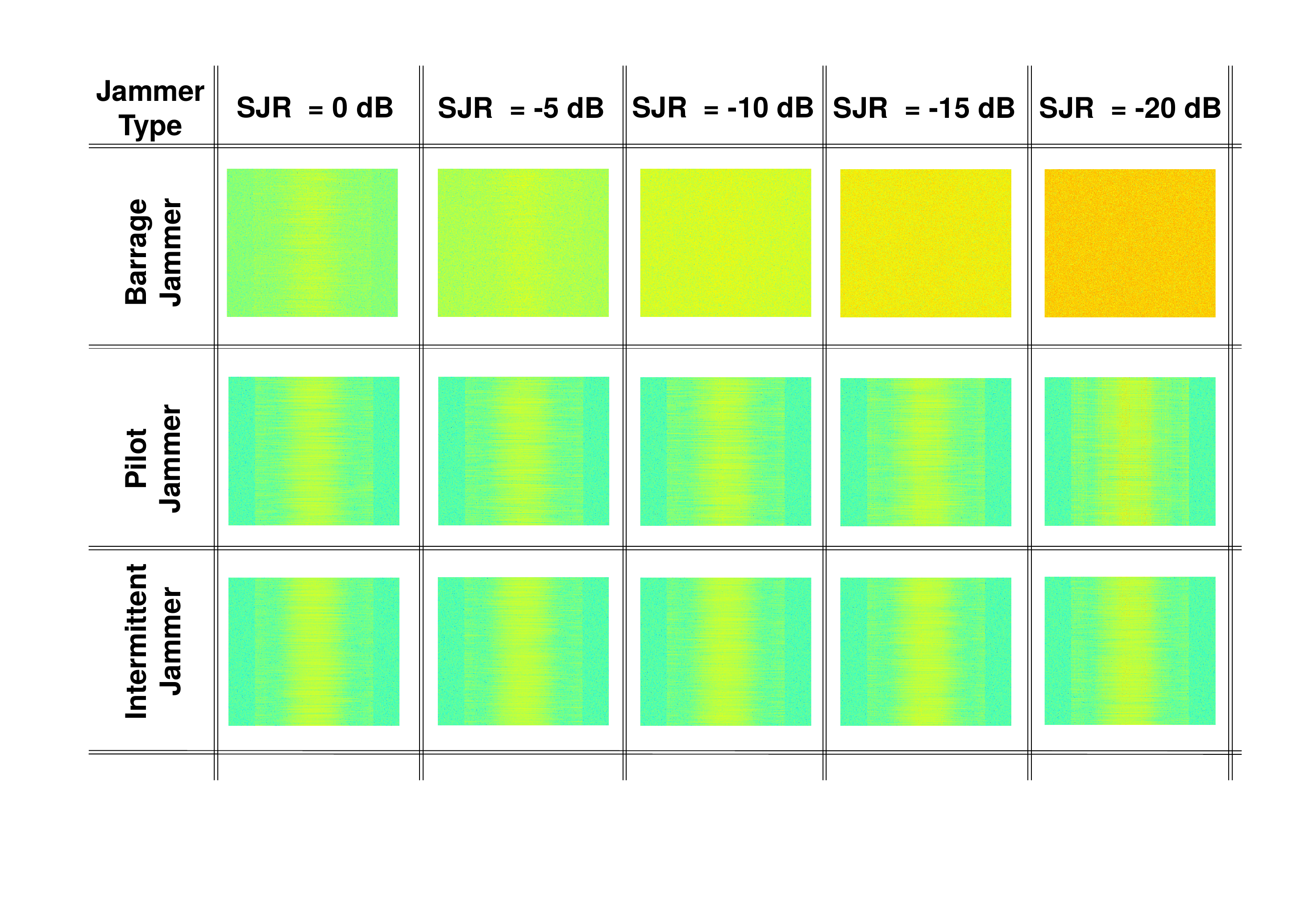}
  \caption{STFT based representations of the received signal under three different jamming attacks when SNR equals 5 dB.}
  \label{fig4}
\end{figure}
\subsection{Benchmark Algorithm: SVM}
The main objective of SVM is the determination of the hyperplane, known as decision boundary, that allows the classification of data samples. The hyperplane is defined as
\begin{equation}
    \mathbf{w}^T\mathbf{x}+b=0,
\end{equation}
where $\mathbf{x}$, $\mathbf{w}$, and $b$ denote the feature vector, support vector, and bias term, respectively. SVM solves the classification problem by maximizing the margin that is the distance between the closest data points and hyperplane. When the data points are not separable linearly, finding optimal hyperplane turns into an optimization problem \cite{datac1}. By solving the optimization problem with the Lagrangian function, the decision function is constructed as
\begin{equation}\label{fx}
    \begin{matrix}
f(x)={\sum_{p=1}^{P} a_{p}  y^{(p)} \mathcal{K}(\textup{\textbf{x}}^{(p)},\textup{\textbf{x}})+b},
\end{matrix}
\end{equation}
where $ y^{(p)} \in \left \{ -1, 1 \right \}$ denotes labels; $P$ is the total number of samples; $a_{p}$ and $\mathcal{K}$ represent the Lagrange multiplier and the kernel function, respectively. According to the decision function output, the classifier predicts the class of data sample.

\begin{figure}[!h]
  \includegraphics[width=0.7\linewidth]{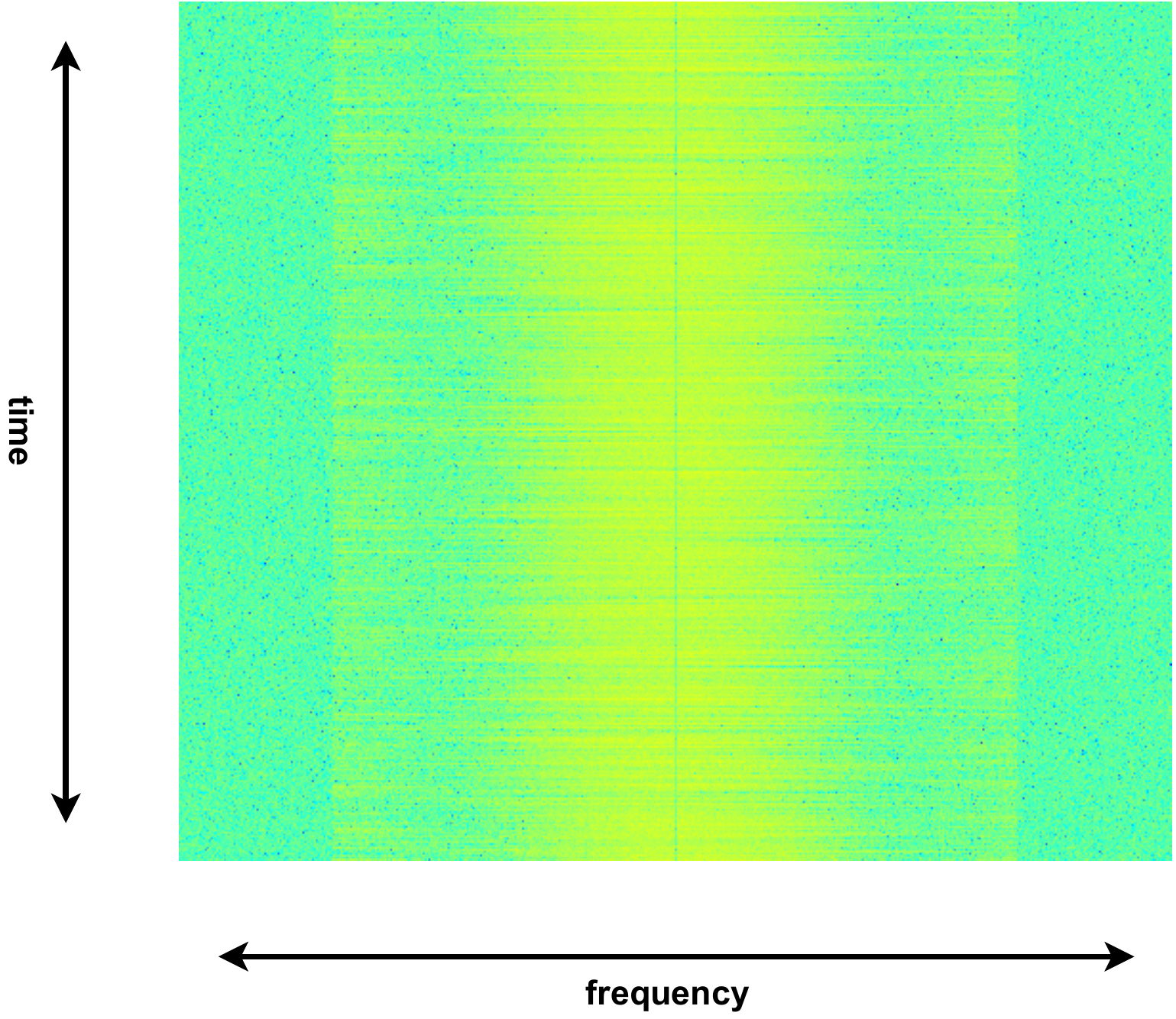}
  \caption{STFT based representations of the received signal under any jamming attacks for 5 dB SNR.}
  \label{fig5}
\end{figure}

\section{Numerical Analysis}\label{IV}

\begin{table}[!h]
\centering
\caption{Parameters.}
\begin{tabular}{|c|l|c|} 
\cline{2-3}
\multicolumn{1}{c|}{}                                                             & \multicolumn{1}{c|}{Parameter} & Value            \\ 
\hline
\multirow{11}{*}{\begin{tabular}[c]{@{}c@{}}Raw Data\\ Generation\end{tabular}}   & Guard time interval            & 64               \\ 
\cline{2-3}
                                                                                  & \# of frames in a data sample  & 10               \\ 
\cline{2-3}
                                                                                  & \# of symbols in a frame       & 60               \\ 
\cline{2-3}
                                                                                  & \# of subcarriers              & 1024             \\ 
\cline{2-3}
                                                                                  & \# of data subcarriers         & 705              \\ 
\cline{2-3}
                                                                                  & $N_p$                             & 88               \\ 
\cline{2-3}
                                                                                  & $l$                              & 8                \\ 
\cline{2-3}
                                                                                  & SNR                            & 5, 10, 15           \\ 
\cline{2-3}
                                                                                  & SJR                            & -20, -15, -10, -5, 0  \\ 
\cline{2-3}
                                                                                  & Modulation type                & BPSK             \\ 
\cline{2-3}
                                                                                  & Rician channel K factor        & 5                \\ 
\hline
\multirow{2}{*}{Pre-Processing}                                                   & FFT size                       & 1024             \\ 
\cline{2-3}
                                                                                  & Sample size                    & 96 x 96          \\ 
\hline
\multirow{3}{*}{\begin{tabular}[c]{@{}c@{}}PCA-SVM\\ Classification\end{tabular}} & PCA component number           & 45               \\ 
\cline{2-3}
                                                                                  & $C$                              & 1                \\ 
\cline{2-3}
                                                                                  & SVM kernel                     & linear           \\ 
\hline
\multirow{7}{*}{CNN}                                                              & Learning rate                  & 0.0001           \\ 
\cline{2-3}
                                                                                  & Kernel size                    & $3\times3$             \\ 
\cline{2-3}
                                                                                  & Kernel stride                  & $1\times1$              \\ 
\cline{2-3}
                                                                                  & $\beta_1$                        & 0.9             \\ 
\cline{2-3}
                                                                                  & $\beta_2$                        & 0.999            \\ 
\cline{2-3}
                                                                                  & Batch size                     & 40               \\ 
\cline{2-3}
                                                                                  & Epoch Number                   & 50               \\
\hline
\end{tabular}\label{table}
\end{table}

\begin{figure}[!t]
  \includegraphics[width=\linewidth]{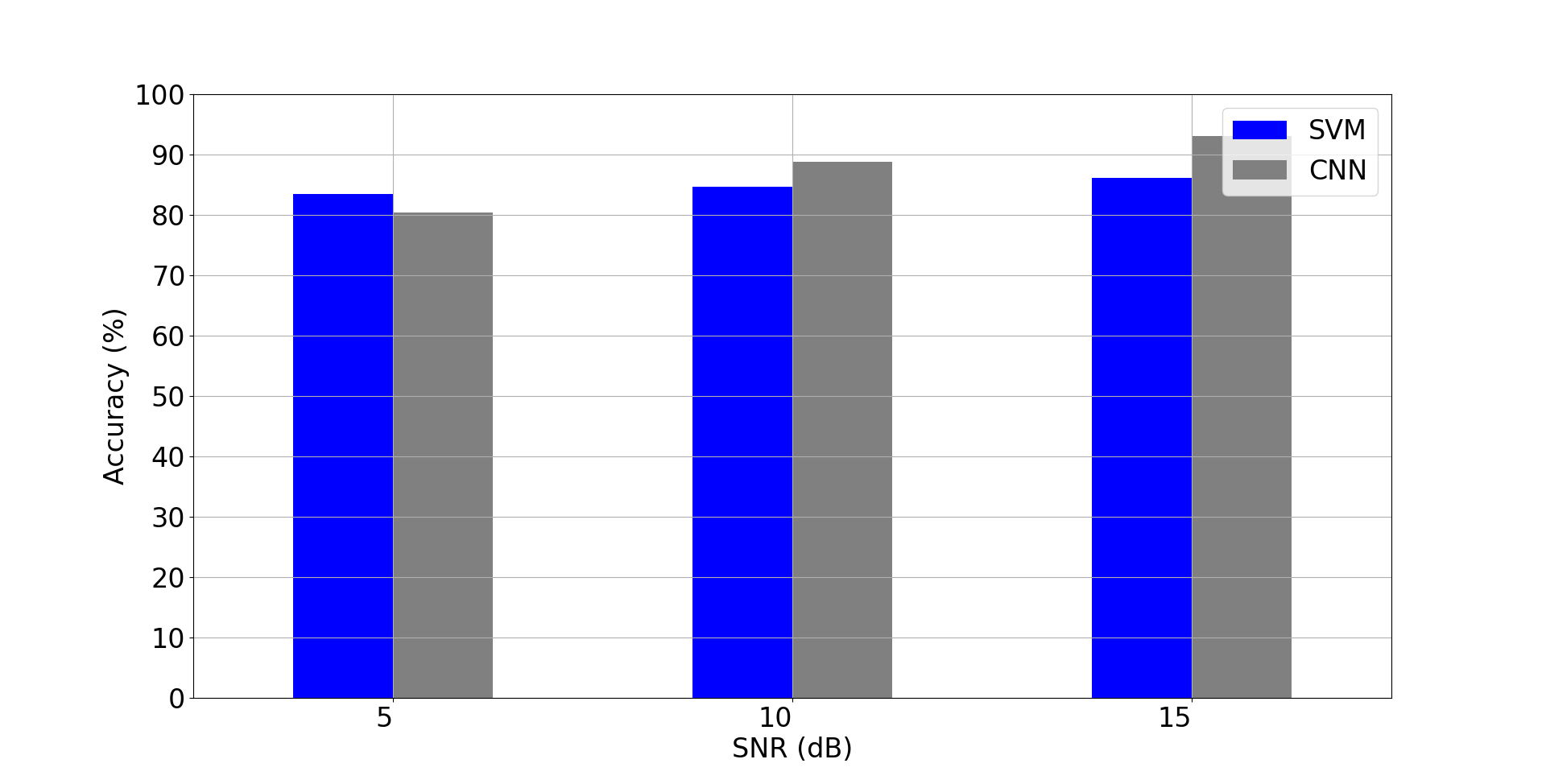}
  \caption{The comparison of SVM and CNN based detection schemes by considering only intermittent jamming attacks are possible.}
  \label{results_inter}
\end{figure}

\begin{figure}[!t]
  \includegraphics[width=\linewidth]{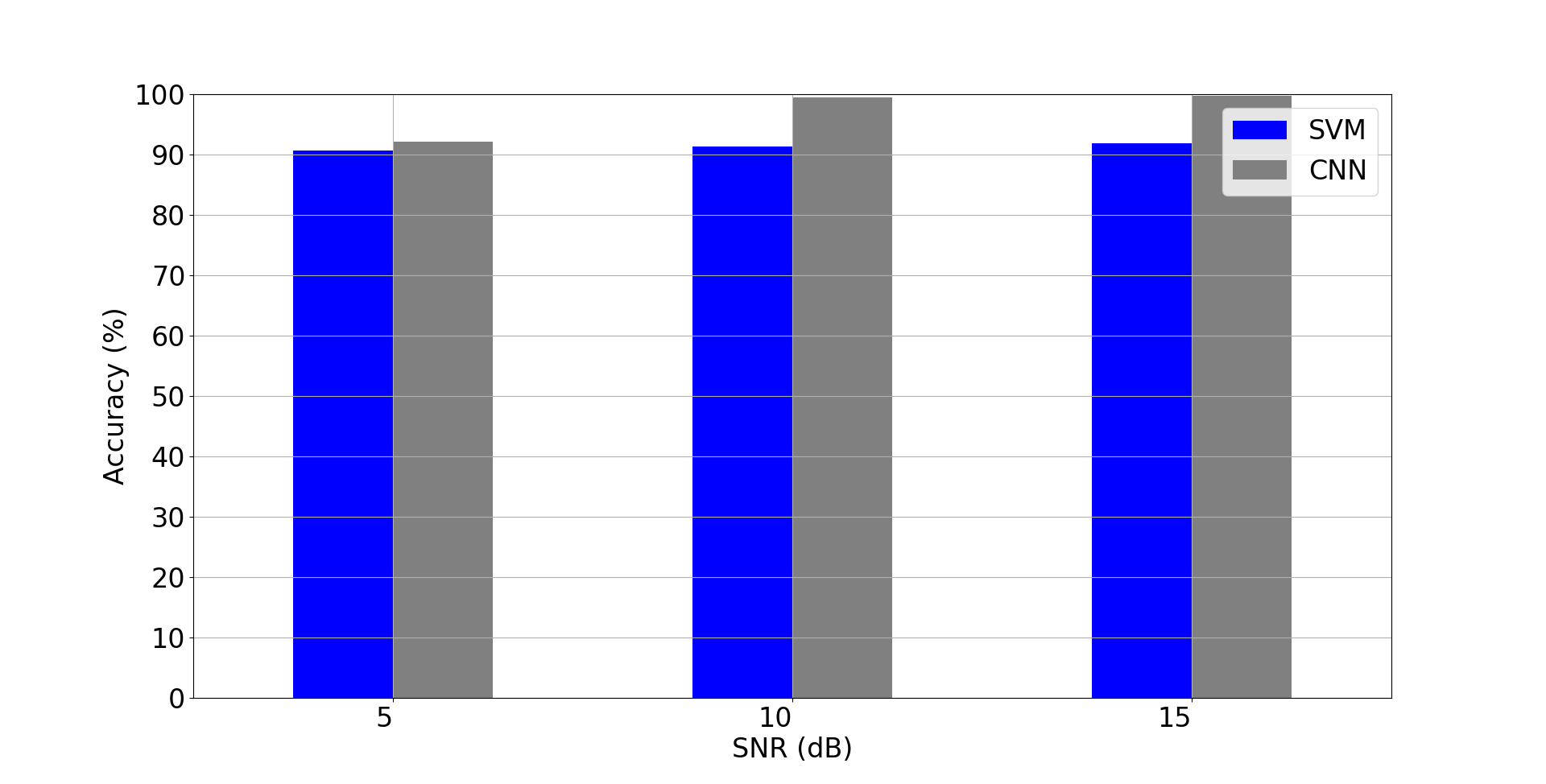}
  \caption{The comparison of SVM and CNN based detection schemes when the all jamming attacks are possible.}
  \label{result_all}
\end{figure}

The telemetry or telecommand signals are generated conceptually following the system model mentioned above with the parameters in Table \ref{table}. We aim to detect the existence of a jamming attack that can be realized with different signal-to-noise ratios (SNR) and signal-to-jamming ratios (SJR). As a solution, learning-driven detection schemes are proposed. A data signal is composed by concatenating $10$ OFDM frames, where each frame includes $60$ OFDM symbols. Short time-frequency transform (STFT) is applied to the received signal as a pre-processing method that improves the detection performance and discriminates the jamming signal characteristic in both frequency and time. The example data samples are demonstrated in Figure \ref{fig4} and Figure \ref{fig5}. 

Firstly, the detection schemes are analyzed under the proposed intermittent jamming strategy that is more critical for satellite communication security due to the importance of these control signals' characteristics. Three datasets are generated under three different SNR levels in Table \ref{table} for training and prediction processes. A dataset is composed of $2000$ training and $3000$ test data samples. The class regarding jammer existence includes data samples in five different SJR levels. Differently, we applied principal component analysis (PCA) in the SVM-based detection scheme for a fair comparison with the CNN-based counterpart, which capable of extracting new features thanks to the convolution operator. The parameters are set as indicated Table \ref{table} during the training processes.

The benchmark results of the detection schemes based on SVM and CNN show high accuracies, as shown in Figure \ref{results_inter}. The SVM-based scheme's accuracies are $83.4\%$, $84.7\%$, and $86.1\%$ for SNR, $5$ dB, $10$ dB, and $15$ dB datasets, respectively. The SVM-based scheme's performance increases slightly depending on SNR values compared to the counterpart. On the other hand, the CNN architecture based scheme demonstrates a considerable increase in detection accuracy where accuracies are $80.4\%$, $88.8\%$, and $93.1\%$ for SNR, $5$ dB, $10$ dB, and $15$ dB datasets, respectively. Additionally, when all SNR values are considered to analyze detectors' performance under a more realistic scenario, three datasets are collected. The proposed CNN architecture based scheme demonstrates superior performance with $96\%$ accuracy than SVM based detector ($84.9\%$ accuracy).

Secondly, we take into account all types of jammer attacks and generate three datasets for each SNR value in Table \ref{table}. Each dataset includes $6000$ training and $9000$ test samples. CNN based detector achieves $92.1\%$, $99.5\%$, and $99.7\%$ when SVM shows $90.6\%$, $91.3\%$, and $91.8\%$ for SNR, $5$ dB, $10$ dB, and $15$ dB, respectively. The results in Figure \ref{result_all} prove that the proposed CNN architecture provides more robust performance in all SNRs than the benchmark algorithm. On the other hand, both schemes demonstrate lower performance to detect the proposed intermittent jamming attack detection, making it critical compared to other attack types.

\section{Conclusion}\label{vv}
We proposed a jamming attack detection scheme based on machine learning by addressing a crucial vulnerability of satellite-based communication systems. The possibility of new threat types is highlighted according to the telemetry and telecommand signaling structure. By considering a conceptual satellite communication system, intermittent jamming attacks are proposed together with common attack types. The presented numerical results prove that these types of attacks can be overcome by employing the proposed detection mechanism.
\bibliographystyle{ACM-Reference-Format}
\bibliography{ref}


\begin{thebibliography}{22}


\ifx \showCODEN    \undefined \def \showCODEN     #1{\unskip}     \fi
\ifx \showDOI      \undefined \def \showDOI       #1{#1}\fi
\ifx \showISBNx    \undefined \def \showISBNx     #1{\unskip}     \fi
\ifx \showISBNxiii \undefined \def \showISBNxiii  #1{\unskip}     \fi
\ifx \showISSN     \undefined \def \showISSN      #1{\unskip}     \fi
\ifx \showLCCN     \undefined \def \showLCCN      #1{\unskip}     \fi
\ifx \shownote     \undefined \def \shownote      #1{#1}          \fi
\ifx \showarticletitle \undefined \def \showarticletitle #1{#1}   \fi
\ifx \showURL      \undefined \def \showURL       {\relax}        \fi
\providecommand\bibfield[2]{#2}
\providecommand\bibinfo[2]{#2}
\providecommand\natexlab[1]{#1}
\providecommand\showeprint[2][]{arXiv:#2}

\bibitem[\protect\citeauthoryear{{Akhlaghpasand}, {Razavizadeh}, {Björnson},
  and {Do}}{{Akhlaghpasand} et~al\mbox{.}}{2018}]%
        {int9}
\bibfield{author}{\bibinfo{person}{H. {Akhlaghpasand}}, \bibinfo{person}{S.~M.
  {Razavizadeh}}, \bibinfo{person}{E. {Björnson}}, {and}
  \bibinfo{person}{T.~T. {Do}}.} \bibinfo{year}{2018}\natexlab{}.
\newblock \bibinfo{title}{Jamming Detection in Massive {MIMO} Systems}.
\newblock , \bibinfo{numpages}{242-245}~pages.
\newblock
\urldef\tempurl%
\url{https://doi.org/10.1109/LWC.2017.2769650}
\showDOI{\tempurl}


\bibitem[\protect\citeauthoryear{{Basar}}{{Basar}}{1983}]%
        {barrage}
\bibfield{author}{\bibinfo{person}{T. {Basar}}.}
  \bibinfo{year}{1983}\natexlab{}.
\newblock \bibinfo{title}{The Gaussian test channel with an intelligent
  jammer}.
\newblock , \bibinfo{numpages}{152-157}~pages.
\newblock
\urldef\tempurl%
\url{https://doi.org/10.1109/TIT.1983.1056602}
\showDOI{\tempurl}


\bibitem[\protect\citeauthoryear{Book}{Book}{2019}]%
        {int5}
\bibfield{author}{\bibinfo{person}{G. Book}.} \bibinfo{year}{2019}\natexlab{}.
\newblock \bibinfo{booktitle}{\emph{The Application of Security to CCCDS
  Protocols}}.
\newblock \bibinfo{type}{{T}echnical {R}eport}.
\newblock


\bibitem[\protect\citeauthoryear{{Cheng}, {Ling}, and {Wu}}{{Cheng}
  et~al\mbox{.}}{2017}]%
        {int12}
\bibfield{author}{\bibinfo{person}{M. {Cheng}}, \bibinfo{person}{Y. {Ling}},
  {and} \bibinfo{person}{W.~B. {Wu}}.} \bibinfo{year}{2017}\natexlab{}.
\newblock \showarticletitle{Time Series Analysis for Jamming Attack Detection
  in Wireless Networks}. In \bibinfo{booktitle}{\emph{IEEE Global
  Communications Conference}}. \bibinfo{pages}{1--7}.
\newblock
\urldef\tempurl%
\url{https://doi.org/10.1109/GLOCOM.2017.8254000}
\showDOI{\tempurl}


\bibitem[\protect\citeauthoryear{Darwish, Kurt, Yanikomeroglu, Lamontagne, and
  Bellemare}{Darwish et~al\mbox{.}}{2021}]%
        {int3}
\bibfield{author}{\bibinfo{person}{Tasneem Darwish}, \bibinfo{person}{Gunes
  Kurt}, \bibinfo{person}{Halim Yanikomeroglu}, \bibinfo{person}{Guillaume
  Lamontagne}, {and} \bibinfo{person}{Michel Bellemare}.}
  \bibinfo{year}{2021}\natexlab{}.
\newblock \bibinfo{title}{Location Management in IP-based Future {LEO}
  Satellite Networks: A Review}.
\newblock
\newblock
\showeprint[arxiv]{cs.NI/2101.08336}


\bibitem[\protect\citeauthoryear{{Di}, {Song}, {Li}, and {Poor}}{{Di}
  et~al\mbox{.}}{2019}]%
        {int4}
\bibfield{author}{\bibinfo{person}{B. {Di}}, \bibinfo{person}{L. {Song}},
  \bibinfo{person}{Y. {Li}}, {and} \bibinfo{person}{H.~V. {Poor}}.}
  \bibinfo{year}{2019}\natexlab{}.
\newblock \bibinfo{title}{Ultra-Dense {LEO}: Integration of Satellite Access
  Networks into 5G and Beyond}.
\newblock , \bibinfo{numpages}{62-69}~pages.
\newblock
\urldef\tempurl%
\url{https://doi.org/10.1109/MWC.2019.1800301}
\showDOI{\tempurl}


\bibitem[\protect\citeauthoryear{{Du}, {Jiang}, {Zhang}, {Wang}, {Ren}, and
  {Debbah}}{{Du} et~al\mbox{.}}{2018}]%
        {int8}
\bibfield{author}{\bibinfo{person}{J. {Du}}, \bibinfo{person}{C. {Jiang}},
  \bibinfo{person}{H. {Zhang}}, \bibinfo{person}{X. {Wang}},
  \bibinfo{person}{Y. {Ren}}, {and} \bibinfo{person}{M. {Debbah}}.}
  \bibinfo{year}{2018}\natexlab{}.
\newblock \bibinfo{title}{Secure Satellite-Terrestrial Transmission Over
  Incumbent Terrestrial Networks via Cooperative Beamforming}.
\newblock , \bibinfo{numpages}{1367-1382}~pages.
\newblock
\urldef\tempurl%
\url{https://doi.org/10.1109/JSAC.2018.2824623}
\showDOI{\tempurl}


\bibitem[\protect\citeauthoryear{{Erpek}, {Sagduyu}, and {Shi}}{{Erpek}
  et~al\mbox{.}}{2019}]%
        {int14}
\bibfield{author}{\bibinfo{person}{T. {Erpek}}, \bibinfo{person}{Y.~E.
  {Sagduyu}}, {and} \bibinfo{person}{Y. {Shi}}.}
  \bibinfo{year}{2019}\natexlab{}.
\newblock \bibinfo{title}{Deep Learning for Launching and Mitigating Wireless
  Jamming Attacks}.
\newblock , \bibinfo{numpages}{2-14}~pages.
\newblock
\urldef\tempurl%
\url{https://doi.org/10.1109/TCCN.2018.2884910}
\showDOI{\tempurl}


\bibitem[\protect\citeauthoryear{Gecgel, Goztepe, and Kurt}{Gecgel
  et~al\mbox{.}}{2019}]%
        {datac2}
\bibfield{author}{\bibinfo{person}{Selen Gecgel}, \bibinfo{person}{Caner
  Goztepe}, {and} \bibinfo{person}{Gunes~Karabulut Kurt}.}
  \bibinfo{year}{2019}\natexlab{}.
\newblock \showarticletitle{Jammer Detection Based on Artificial Neural
  Networks: A Measurement Study}. In \bibinfo{booktitle}{\emph{Proceedings of
  the ACM Workshop on Wireless Security and Machine Learning}}
  \emph{(\bibinfo{series}{WiseML 2019})}. \bibinfo{publisher}{Association for
  Computing Machinery}, \bibinfo{address}{New York, NY, USA},
  \bibinfo{pages}{43–48}.
\newblock
\showISBNx{9781450367691}
\urldef\tempurl%
\url{https://doi.org/10.1145/3324921.3328788}
\showDOI{\tempurl}


\bibitem[\protect\citeauthoryear{{Guo}, {An}, {Zhang}, {Huang}, {Tang},
  {Zheng}, and {Tsiftsis}}{{Guo} et~al\mbox{.}}{2020}]%
        {int7}
\bibfield{author}{\bibinfo{person}{K. {Guo}}, \bibinfo{person}{K. {An}},
  \bibinfo{person}{B. {Zhang}}, \bibinfo{person}{Y. {Huang}},
  \bibinfo{person}{X. {Tang}}, \bibinfo{person}{G. {Zheng}}, {and}
  \bibinfo{person}{T.~A. {Tsiftsis}}.} \bibinfo{year}{2020}\natexlab{}.
\newblock \bibinfo{title}{Physical Layer Security for Multiuser Satellite
  Communication Systems With Threshold-Based Scheduling Scheme}.
\newblock , \bibinfo{numpages}{5129-5141}~pages.
\newblock
\urldef\tempurl%
\url{https://doi.org/10.1109/TVT.2020.2979496}
\showDOI{\tempurl}


\bibitem[\protect\citeauthoryear{Harel}{Harel}{2008}]%
        {int6}
\bibfield{author}{\bibinfo{person}{D. Harel}.} \bibinfo{year}{2008}\natexlab{}.
\newblock \bibinfo{booktitle}{\emph{Satellite Earth Stations and Systems (SES);
  Satellite Component of UMTS/IMT-2000; Evaluation of the OFDM as a Satellite
  Radio Interface}}.
\newblock \bibinfo{type}{{T}echnical {R}eport}.
\newblock


\bibitem[\protect\citeauthoryear{{Jahanshahi} and {Eslami}}{{Jahanshahi} and
  {Eslami}}{2011}]%
        {int13}
\bibfield{author}{\bibinfo{person}{J.~A. {Jahanshahi}} {and}
  \bibinfo{person}{M. {Eslami}}.} \bibinfo{year}{2011}\natexlab{}.
\newblock \showarticletitle{On the performance of SVM based jamming attacks
  detection algorithm in base station}. In \bibinfo{booktitle}{\emph{IEEE
  Swedish Communication Technologies Workshop (Swe-CTW)}}.
  \bibinfo{pages}{109--113}.
\newblock
\urldef\tempurl%
\url{https://doi.org/10.1109/Swe-CTW.2011.6082476}
\showDOI{\tempurl}


\bibitem[\protect\citeauthoryear{Kingma and Ba}{Kingma and Ba}{2015}]%
        {ADAM}
\bibfield{author}{\bibinfo{person}{Diederick~P Kingma} {and}
  \bibinfo{person}{Jimmy Ba}.} \bibinfo{year}{2015}\natexlab{}.
\newblock \showarticletitle{ADAM: A method for stochastic optimization}. In
  \bibinfo{booktitle}{\emph{Inter. Conf. on Learning Rep. (ICLR)}}.
\newblock


\bibitem[\protect\citeauthoryear{{Papathanassiou}, {Salkintzis}, and
  {Mathiopoulos}}{{Papathanassiou} et~al\mbox{.}}{2001}]%
        {OFDM}
\bibfield{author}{\bibinfo{person}{A. {Papathanassiou}}, \bibinfo{person}{A.~K.
  {Salkintzis}}, {and} \bibinfo{person}{P.~T. {Mathiopoulos}}.}
  \bibinfo{year}{2001}\natexlab{}.
\newblock \bibinfo{title}{A comparison study of the uplink performance of
  {W-CDMA} and {OFDM} for mobile multimedia communications via {LEO}
  satellites}.
\newblock , \bibinfo{numpages}{35-43}~pages.
\newblock
\urldef\tempurl%
\url{https://doi.org/10.1109/98.930095}
\showDOI{\tempurl}


\bibitem[\protect\citeauthoryear{Pratt and Allnutt}{Pratt and Allnutt}{2019}]%
        {int11}
\bibfield{author}{\bibinfo{person}{T. Pratt} {and} \bibinfo{person}{J.E.
  Allnutt}.} \bibinfo{year}{2019}\natexlab{}.
\newblock \bibinfo{booktitle}{\emph{Satellite Communications}}.
\newblock \bibinfo{publisher}{John Wiley \& Sons}.
\newblock
\showISBNx{9781119482178}
\showLCCN{2019015618}


\bibitem[\protect\citeauthoryear{{Qu}, {Zhang}, {Cao}, and {Xie}}{{Qu}
  et~al\mbox{.}}{2017}]%
        {int1}
\bibfield{author}{\bibinfo{person}{Z. {Qu}}, \bibinfo{person}{G. {Zhang}},
  \bibinfo{person}{H. {Cao}}, {and} \bibinfo{person}{J. {Xie}}.}
  \bibinfo{year}{2017}\natexlab{}.
\newblock \bibinfo{title}{{LEO} Satellite Constellation for Internet of
  Things}.
\newblock , \bibinfo{numpages}{18391-18401}~pages.
\newblock
\urldef\tempurl%
\url{https://doi.org/10.1109/ACCESS.2017.2735988}
\showDOI{\tempurl}


\bibitem[\protect\citeauthoryear{{Shi}, {Davaslioglu}, {Sagduyu}, {Headley},
  {Fowler}, and {Green}}{{Shi} et~al\mbox{.}}{2019}]%
        {shi1}
\bibfield{author}{\bibinfo{person}{Y. {Shi}}, \bibinfo{person}{K.
  {Davaslioglu}}, \bibinfo{person}{Y.~E. {Sagduyu}}, \bibinfo{person}{W.~C.
  {Headley}}, \bibinfo{person}{M. {Fowler}}, {and} \bibinfo{person}{G.
  {Green}}.} \bibinfo{year}{2019}\natexlab{}.
\newblock \showarticletitle{Deep Learning for RF Signal Classification in
  Unknown and Dynamic Spectrum Environments}. In \bibinfo{booktitle}{\emph{IEEE
  International Symposium on Dynamic Spectrum Access Networks (DySPAN)}}.
  \bibinfo{pages}{1--10}.
\newblock
\urldef\tempurl%
\url{https://doi.org/10.1109/DySPAN.2019.8935684}
\showDOI{\tempurl}


\bibitem[\protect\citeauthoryear{Shi, Sagduyu, Erpek, Davaslioglu, Lu, and
  Li}{Shi et~al\mbox{.}}{2018}]%
        {shi2}
\bibfield{author}{\bibinfo{person}{Y. Shi}, \bibinfo{person}{Yalin~E. Sagduyu},
  \bibinfo{person}{T. Erpek}, \bibinfo{person}{Kemal Davaslioglu},
  \bibinfo{person}{Zhuo Lu}, {and} \bibinfo{person}{J. Li}.}
  \bibinfo{year}{2018}\natexlab{}.
\newblock \showarticletitle{Adversarial Deep Learning for Cognitive Radio
  Security: Jamming Attack and Defense Strategies}.
\newblock \bibinfo{journal}{\emph{IEEE International Conference on
  Communications Workshops (ICC Workshops)}} (\bibinfo{year}{2018}),
  \bibinfo{pages}{1--6}.
\newblock


\bibitem[\protect\citeauthoryear{{Su}, {Liu}, {Zhou}, {Yuan}, {Cao}, and
  {Shi}}{{Su} et~al\mbox{.}}{2019}]%
        {int2}
\bibfield{author}{\bibinfo{person}{Y. {Su}}, \bibinfo{person}{Y. {Liu}},
  \bibinfo{person}{Y. {Zhou}}, \bibinfo{person}{J. {Yuan}}, \bibinfo{person}{H.
  {Cao}}, {and} \bibinfo{person}{J. {Shi}}.} \bibinfo{year}{2019}\natexlab{}.
\newblock \bibinfo{title}{Broadband {LEO} Satellite Communications:
  Architectures and Key Technologies}.
\newblock , \bibinfo{numpages}{55-61}~pages.
\newblock
\urldef\tempurl%
\url{https://doi.org/10.1109/MWC.2019.1800299}
\showDOI{\tempurl}


\bibitem[\protect\citeauthoryear{Topal, Gecgel, Eksioglu, and {Karabulut
  Kurt}}{Topal et~al\mbox{.}}{2020}]%
        {datac1}
\bibfield{author}{\bibinfo{person}{Ozan~Alp Topal}, \bibinfo{person}{Selen
  Gecgel}, \bibinfo{person}{Ender~Mete Eksioglu}, {and} \bibinfo{person}{Gunes
  {Karabulut Kurt}}.} \bibinfo{year}{2020}\natexlab{}.
\newblock \bibinfo{title}{Identification of smart jammers: Learning-based
  approaches using wavelet preprocessing}.
\newblock , \bibinfo{numpages}{101029}~pages.
\newblock
\showISSN{1874-4907}
\urldef\tempurl%
\url{https://doi.org/10.1016/j.phycom.2020.101029}
\showDOI{\tempurl}


\bibitem[\protect\citeauthoryear{{Xu}, {Xu}, {Pan}, and {Elkashlan}}{{Xu}
  et~al\mbox{.}}{2019}]%
        {int10}
\bibfield{author}{\bibinfo{person}{S. {Xu}}, \bibinfo{person}{W. {Xu}},
  \bibinfo{person}{C. {Pan}}, {and} \bibinfo{person}{M. {Elkashlan}}.}
  \bibinfo{year}{2019}\natexlab{}.
\newblock \bibinfo{title}{Detection of Jamming Attack in Non-Coherent Massive
  {SIMO} Systems}.
\newblock , \bibinfo{numpages}{2387-2399}~pages.
\newblock
\urldef\tempurl%
\url{https://doi.org/10.1109/TIFS.2019.2899484}
\showDOI{\tempurl}


\bibitem[\protect\citeauthoryear{{You}, {Li}, {Wang}, {Gao}, {Xia}, and
  {Ottersten}}{{You} et~al\mbox{.}}{2020}]%
        {Rician}
\bibfield{author}{\bibinfo{person}{L. {You}}, \bibinfo{person}{K.~X. {Li}},
  \bibinfo{person}{J. {Wang}}, \bibinfo{person}{X. {Gao}},
  \bibinfo{person}{X.~G. {Xia}}, {and} \bibinfo{person}{B. {Ottersten}}.}
  \bibinfo{year}{2020}\natexlab{}.
\newblock \bibinfo{title}{Massive MIMO Transmission for LEO Satellite
  Communications}.
\newblock , \bibinfo{numpages}{1851-1865}~pages.
\newblock
\urldef\tempurl%
\url{https://doi.org/10.1109/JSAC.2020.3000803}
\showDOI{\tempurl}


\end{thebibliography}
\end{document}